\documentclass[%
reprint,
superscriptaddress,
amsmath,amssymb,
aps,
noeprint,
]{revtex4-2}

\usepackage{amsmath}
\usepackage{color}
\usepackage[utf8]{inputenc}
\usepackage{graphicx} 
\usepackage{dcolumn} 
\usepackage{bm} 
\usepackage{upgreek}
\usepackage[exponent-product=\cdot]{siunitx}
\usepackage{eqnarray}
\usepackage{multirow}
\usepackage{hyperref}

\begin{document}

\title{New Constraint on the Local Relic Neutrino Background Overdensity with the First KATRIN Data Runs}

\newcommand{\correspondingauthors}{Corresponding authors: \\ kellerer@mpp.mpg.de and  thierry.lasserre@cea.fr}
\newcommand{\berlin}{Institut f\"{u}r Physik, Humboldt-Universit\"{a}t zu Berlin, Newtonstr. 15, 12489 Berlin, Germany}
\newcommand{\bonn}{Helmholtz-Institut f\"{u}r Strahlen- und Kernphysik, Rheinische Friedrich-Wilhelms-Universit\"{a}t Bonn, Nussallee 14-16, 53115 Bonn, Germany}
\newcommand{\cmu}{Department of Physics, Carnegie Mellon University, Pittsburgh, PA 15213, USA}
\newcommand{\cwru}{Department of Physics, Case Western Reserve University, Cleveland, OH 44106, USA}
\newcommand{\etp}{Institute of Experimental Particle Physics~(ETP), Karlsruhe Institute of Technology~(KIT), Wolfgang-Gaede-Str. 1, 76131 Karlsruhe, Germany}
\newcommand{\fulda}{University of Applied Sciences~(HFD)~Fulda, Leipziger Str.~123, 36037 Fulda, Germany}
%
%%% BEGIN: KIT institutes
\newcommand{\iap}{Institute for Astroparticle Physics~(IAP), Karlsruhe Institute of Technology~(KIT), Hermann-von-Helmholtz-Platz 1, 76344 Eggenstein-Leopoldshafen, Germany}
\newcommand{\ipe}{Institute for Data Processing and Electronics~(IPE), Karlsruhe Institute of Technology~(KIT), Hermann-von-Helmholtz-Platz 1, 76344 Eggenstein-Leopoldshafen, Germany}
\newcommand{\itep}{Institute for Technical Physics~(ITEP), Karlsruhe Institute of Technology~(KIT), Hermann-von-Helmholtz-Platz 1, 76344 Eggenstein-Leopoldshafen, Germany}
\newcommand{\tlk}{Tritium Laboratory Karlsruhe~(TLK), Karlsruhe Institute of Technology~(KIT), Hermann-von-Helmholtz-Platz 1, 76344 Eggenstein-Leopoldshafen, Germany}
\newcommand{\ppq}{Project, Process, and Quality Management~(PPQ), Karlsruhe Institute of Technology~(KIT), Hermann-von-Helmholtz-Platz 1, 76344 Eggenstein-Leopoldshafen, Germany    }
%
%%% END: KIT Institutes
%
\newcommand{\inr}{Institute for Nuclear Research of Russian Academy of Sciences, 60th October Anniversary Prospect 7a, 117312 Moscow, Russia}
\newcommand{\lbnl}{Institute for Nuclear and Particle Astrophysics and Nuclear Science Division, Lawrence Berkeley National Laboratory, Berkeley, CA 94720, USA}
\newcommand{\madrid}{Departamento de Qu\'{i}mica F\'{i}sica Aplicada, Universidad Autonoma de Madrid, Campus de Cantoblanco, 28049 Madrid, Spain}
\newcommand{\mainz}{Institut f\"{u}r Physik, Johannes-Gutenberg-Universit\"{a}t Mainz, 55099 Mainz, Germany}
\newcommand{\mpp}{Max-Planck-Institut f\"{u}r Physik, F\"{o}hringer Ring 6, 80805 M\"{u}nchen, Germany}
\newcommand{\massit}{Laboratory for Nuclear Science, Massachusetts Institute of Technology, 77 Massachusetts Ave, Cambridge, MA 02139, USA}
\newcommand{\mpik}{Max-Planck-Institut f\"{u}r Kernphysik, Saupfercheckweg 1, 69117 Heidelberg, Germany}
\newcommand{\muenster}{Institute for Nuclear Physics, University of M\"{u}nster, Wilhelm-Klemm-Str. 9, 48149 M\"{u}nster, Germany}
\newcommand{\npi}{Nuclear Physics Institute,  Czech Academy of Sciences, 25068 \v{R}e\v{z}, Czech Republic}
\newcommand{\unc}{Department of Physics and Astronomy, University of North Carolina, Chapel Hill, NC 27599, USA}
\newcommand{\washington}{Center for Experimental Nuclear Physics and Astrophysics, and Dept.~of Physics, University of Washington, Seattle, WA 98195, USA}
\newcommand{\wuppertal}{Department of Physics, Faculty of Mathematics and Natural Sciences, University of Wuppertal, Gau{\ss}str. 20, 42119 Wuppertal, Germany}
\newcommand{\saclay}{IRFU (DPhP \& APC), CEA, Universit\'{e} Paris-Saclay, 91191 Gif-sur-Yvette, France }
\newcommand{\tum}{Technische Universit\"{a}t M\"{u}nchen, James-Franck-Str. 1, 85748 Garching, Germany}
\newcommand{\uhd}{Institute for Theoretical Astrophysics, University of Heidelberg, Albert-Ueberle-Str. 2, 69120 Heidelberg, Germany}
\newcommand{\tunl}{Triangle Universities Nuclear Laboratory, Durham, NC 27708, USA}
%
%%% BEGIN: other institutions
%
\newcommand{\ornl}{Also affiliated with Oak Ridge National Laboratory, Oak Ridge, TN 37831, USA}
%
%\newcommand{\swansea}{Department of Physics, Swansea University, Singleton Park, Swansea SA2 8PP, United Kingdom}
%\newcommand{\ucsb}{Department of Physics, University of California at Santa Barbara, Santa Barbara, CA 93106, USA}
%
%%% END: other institutions
%

\affiliation{\tlk}
\affiliation{\ipe}
\affiliation{\iap}
\affiliation{\inr}
\affiliation{\muenster}
\affiliation{\etp}
\affiliation{\itep}
\affiliation{\tum}
\affiliation{\mpp}
\affiliation{\unc}
\affiliation{\tunl}
\affiliation{\lbnl}
\affiliation{\wuppertal}
\affiliation{\madrid}
\affiliation{\washington}
\affiliation{\npi}
\affiliation{\massit}
\affiliation{\cmu}
\affiliation{\saclay}
\affiliation{\mpik}
\affiliation{\berlin}
\affiliation{\uhd}
\affiliation{\mainz}

%\affiliation{\fulda}
%\affiliation{\bonn}
%\affiliation{\cwru}
%\affiliation{\ppq}

\author{M.~Aker}\affiliation{\tlk}
\author{D.~Batzler}\affiliation{\tlk}
\author{A.~Beglarian}\affiliation{\ipe}
\author{J.~Behrens}\affiliation{\iap}
\author{A.~Berlev}\affiliation{\inr}
\author{U.~Besserer}\affiliation{\tlk}
\author{B.~Bieringer}\affiliation{\muenster}
\author{F.~Block}\affiliation{\etp}
\author{S.~Bobien}\affiliation{\itep}
\author{B.~Bornschein}\affiliation{\tlk}
% ...

\collaboration{KATRIN Collaboration}\noaffiliation

\date{\today}

\begin{abstract} 
 We report on the direct cosmic relic neutrino background search from the first two science runs of the KATRIN experiment in 2019. Beta-decay electrons from a high-purity molecular tritium gas source are analyzed by a high-resolution MAC-E filter around the kinematic endpoint at \SI{18.57}{keV}. The analysis is sensitive to a local relic neutrino overdensity of $ \eta$ = \num{9.7e10} (\num{1.1e11}) at a 90\% (95\%) confidence level. A fit of the integrated electron spectrum over a narrow interval around the kinematic endpoint accounting for relic neutrino captures in the tritium source reveals no significant overdensity. This work improves the results obtained by the previous kinematic neutrino mass experiments at Los Alamos and Troitsk. We furthermore update the projected final sensitivity of the KATRIN experiment to $ \eta < \num{1e10} $ at 90\% confidence level, by relying on updated operational conditions.
\end{abstract}

\keywords{Suggested keywords}
\maketitle

\begin{figure}[ht!]
		\includegraphics[width=0.45\textwidth]{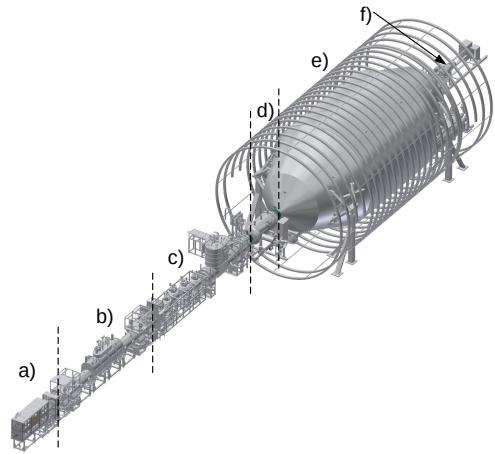}
		\caption{\label{fig:katrin_setup} The main components of KATRIN: a) the rear Section, b) the windowless
			gaseous tritium source WGTS, c) the pumping section, and a tandem set-up of two MAC-E-
			filters: d) the smaller pre-spectrometer and e) the larger main spectrometer with surrounding aircoil system. This setup allows only the highest energy $ \beta$-decay electrons to reach the focal plane detector f) where they are counted.
		}
\end{figure}

{\em Introduction} -- In modern cosmology, neutrinos decoupled from the other particles of the standard model when the universe was about one second old. The existence of a cosmic (or relic) neutrino background (C$\nu$B) is predicted with great confidence. A direct measurement of C$\nu$B remains one of the most difficult tasks in neutrino physics and would yield direct information about the early history of the universe. Theoretical predictions of the local relic neutrino overdensity ($\eta$) range from 1.2 to 20 for neutrino masses below \SI{0.6}{\electronvolt}, depending on the neutrino mass and the assumed density profile of the Milky Way \cite{Ringwald_2005,de_Salas_2017}. Relic neutrinos can interact with radioactive nuclei, like tritium, via the neutrino capture reaction $ \nu_e+N^A_Z\longrightarrow N^A_{Z+1}+e^-$ \cite{PhysRev.128.1457,Faessler_2017}. On this basis, previous kinematic neutrino mass experiments have provided upper bounds on the local overdensity $ \eta $ of relic neutrinos (with respect to the average across the universe of \SI{56}{cm^{-3}} per species \cite{Ringwald_2005}) of \num{1e13} \cite{LOBASHEV1999327,Robertson:1991vn}. Using a similar but extensively improved technology, the Karlsruhe Tritium Neutrino Experiment (KATRIN), shown in figure~\ref{fig:katrin_setup}, provides a high-precision measurement of the electron spectrum of the tritium $\beta$-decay,  $^3\mathrm{H}\rightarrow~^3\mathrm{He}^+ + \mathrm{e}^- + \overline{\nu}_e $ (endpoint $ E_0 $ = \SI{18.57}{keV}, half-life $ t_{1/2} $ = \SI{12.32}{yr}). Primarily operated to measure the effective neutrino mass $ m_\nu $, KATRIN already holds an upper limit of $ m_\nu < $ \SI{0.8}{eV} at 90\% confidence level \cite{aker2021direct} following the first two science runs in 2019. Using the same data set \cite{KNM1_PRD,Aker_2021,aker2021direct}, we establish new constraints on the local overdensity of relic neutrinos. 

{\em Experimental setup} -- KATRIN combines a windowless gaseous molecular tritium source (WGTS) \cite{Robertson:1991vn}, with two spectrometers based on the principle of magnetic adiabatic collimation with electrostatic filtering (MAC-E-filter) \cite{Lobashev:1985mu,Picard1992,Kraus2005,Aseev2011}. Figure~\ref{fig:katrin_setup} displays the \SI{70}{m} long experimental setup located at the Karlsruhe Institute of Technology (KIT) in Germany.
%KRN paper KATRIN figure.pdf

 Source-related components in contact with tritium, shown in figure~\ref{fig:katrin_setup} a-c, are part of the Tritium Laboratory Karlsruhe to enable a closed cycle of tritium \cite{Priester:2015bfa}. High-purity tritium gas is continuously injected into the WGTS at \SI{30}{K}. The gas diffuses from the center to the ends of the WGTS where it is pumped out by several turbomolecular pumps and a cryotrap in the pumping section (figure~\ref{fig:katrin_setup} c). This reduces the flow rate of tritium into the spectrometer and detector sections (figure~\ref{fig:katrin_setup} d-f) by more than 14 orders of magnitude to suppress source-related background \cite{Osipowicz:2001sq,Arenz:2018jpa}. 
%This is further facilitated by the \SI{3}{K} cryotrap (CPS), through which the electrons are adiabatically guided towards the spectrometers by the source magnetic field ($ B_\mathrm{WGTS} = \SI{2.52}{T} $) and other superconducting magnets \cite{Arenz:2018jpa}. 
In the spectrometer section, using the MAC-E-filter technique, electrons of charge $ q $ are guided by the magnetic field and precisely filtered by an electrostatic barrier (retarding energy $ qU $). Only electrons whose energy is sufficient to overcome this barrier are transmitted to the focal plane detector (figure~\ref{fig:katrin_setup} f). By varying the High Voltage (HV) setting $ U $, the tritium $ \beta $-decay spectrum is measured in an integral mode, with an energy resolution of $ \Delta E = \SI{2.8}{\electronvolt} $ at $ E_\mathrm{0} $. Transmitted electrons are subsequently counted, as a function of $ qU $, by the focal plane silicon detector, segmented in 148 pixels \cite{Amsbaugh:2014uca}. Details and performance of the KATRIN setup, specified in \cite{KDR2004}, are reported in \cite{Arenz:2018kma,Priester:2015bfa,Arenz:2018jpa,Arenz:2018ymp}.

\begin{figure}[ht!]
	\includegraphics[width=0.45\textwidth]{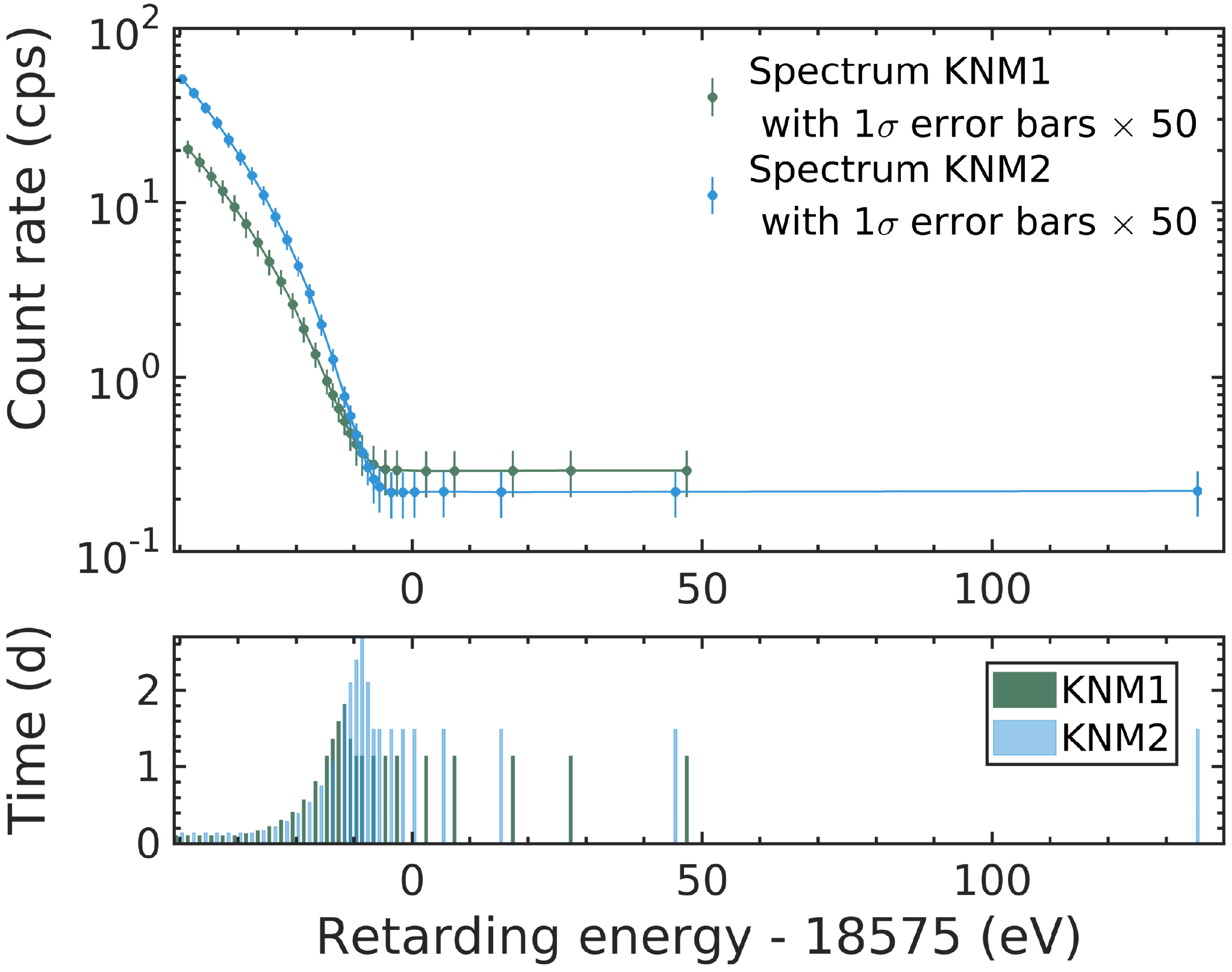}
	\caption{\label{fig:DataSets} Uniform electron spectra $ R(\langle qU\rangle) $ of both data sets, with 1-$ \sigma $ error enlarged by factor 50. The blue and gray lines represent the $ \beta $-decay best-fit models $ R_\mathrm{\beta,calc}(\langle qU \rangle) $. Above $ E_\mathrm{0} $, we see a flat background $ R_\mathrm{bg} $ (top). Integral measurement time distributions of both data sets (bottom).
	}
\end{figure}

{\em First measurement campaign (KNM1)} --  The first science run was carried out from April 10 to May 13, 2019. All experimental details were already reported in \cite{KNM1_PRD}. The average source activity was \SI{2.45e10}{Bq} (\SI{3.4}{\micro\gram} of tritium) at a column density of \num{1.11e17} molecules $ \mathrm{cm}^{-2} $, corresponding to  about 20\% of its nominal value. The integral tritium $ \beta $-decay spectrum was scanned continuously in a range of [$ E_\mathrm{0} $-\SI{91}{\electronvolt}, $ E_\mathrm{0} $+\SI{49}{\electronvolt}] over a series of non-equidistant HV settings of the spectrometer electrode system. %A single scan takes a net time of \SI{2}{h}.
At each HV set point, the number of transmitted electrons is measured.
%in time intervals ranging in length from 17 to 576 s. 
In this search for relic neutrinos, we analyze the region from \SI{37}{\electronvolt} below $ E_\mathrm{0} $ (22 HV set points) to \SI{49}{\electronvolt} above (5 HV set points), as shown in the measurement time distribution (MTD) in figure~\ref{fig:DataSets} (bottom). We first merge the 117 best detector pixels into a single unique effective pixel. Quality cuts in the slow control parameters associated with each tritium scan shorten our data set to 274 stable scans for an overall scan time of \SI{521.7}{h}. The good timing stability and reproducibility of the HV set points allowed us to stack the data of the 274 scans into a single spectrum, which is displayed in figure~\ref{fig:DataSets} (top) in units of counts per second (cps) \cite{KNM1_PRD}. This stacked integral spectrum, $ R(\langle qU\rangle) $, includes \num{1.48e6} $ \beta $-decay electrons expected below $ E_\mathrm{0} $. The background (292 mcps) originates mainly from the spectrometer and has two primary sources.  A significant part is contributed by the thermal ionization of Rydberg atoms that sputter off the inner spectrometer surfaces by $ ^{206} $Pb-recoil ions following $ \alpha $-decays of $ ^{210} $Po. Another source of background are secondary electrons induced by $ \alpha $-decays of single $ ^{219} $Rn atoms emanating from the vacuum pumps of the large spectrometer \cite{FRANKLE2011128}. These electrons start at sub-eV energies, but are later accelerated to $ qU $ by the MAC-E-filter. The radon-induced background also causes a small non-Poissonian rate over-dispersion increasing the background statistical uncertainty (see \cite{KNM1_PRD} for details).

%DataSets.pdf

{\em Second measurement campaign (KNM2)} -- The second measurement campaign was conducted from September 27 to November 19, 2019. The tritium source was operated at its nominal activity of $ \SI{9.5e10}{Bq}$ (\SI{13.0}{\micro\gram} of tritium), yielding more than twice the number of $ \beta$-decay electrons in the analysis range from $ E_0 $-\SI{40}{\electronvolt} to $ E_0 $ at \num{3.68e6} \cite{aker2021direct} as for the KNM1 cycle. To better assess the background, the analysis interval was extended to [$ E_\mathrm{0} $-40 eV, $ E_\mathrm{0} $+135 eV]. Improved vacuum conditions enabled a 25\% reduction of the spectrometer background compared to KNM1  \cite{Arenz:2016mrh}. The KNM2 science run led to the collection of a data set of 361 stable scans with an overall scan time of \SI{743.7}{h} that were combined into an effective single spectrum after gathering the 117 best pixels \cite{aker2021direct}. \newline

{\em Modeling} -- The modeled experimental spectrum $ R_\mathrm{calc}(\langle qU\rangle) $ is calculated by convolving the expected differential spectrum of the signal $ R_\mathrm{diff}(E) $ with the calculated response function $ f(E-\langle qU\rangle) $, in addition to the qU-independent background rate $ R_\mathrm{bg} $: 
\begin{multline}
	R_\mathrm{calc}(\langle qU\rangle) = \int_{qU}^{E_0} R_\mathrm{diff}(E)~\cdot~f(E-\langle qU\rangle)~dE \\
	+ R_\mathrm{bg}
\end{multline}
The response function $ f(E-\langle qU\rangle) $ gives the transmission probability of an electron as a function of its surplus energy $ E-\langle qU\rangle $. It includes the angular spread of electrons and their probability to undergo inelastic scattering processes (see \cite{KNM1_PRD} for details).
Accounting for the relic neutrinos, the differential spectrum $ R_\mathrm{diff}(E) $ is the sum of the $ \beta $-decay and neutrino-capture differential spectra, $ R_{\beta}(E) $ and $ R_{\mathrm{C\nu B}}(E) $ respectively:
\begin{equation}
R_\mathrm{diff}(E)=R_{\beta}(E)+R_{\mathrm{C\nu B}}(E) \, .
\end{equation}
The differential spectrum $ R_\beta(E) $ from the super-allowed $ \beta $-decay of molecular tritium is given by:
\begin{multline}
R_\beta(E)=A_\mathrm{s}\cdot N_\mathrm{T}\cdot\frac{G_\mathrm{F}^2\cdot\mathrm{cos}^2\Theta_\mathrm{C}}{2\pi^3}\cdot |M^2_\mathrm{nucl}| \\
\cdot F(E,Z')\cdot (E+m_\mathrm{e})\cdot\sqrt{(E+m_\mathrm{e})^2-m_\mathrm{e}^2} \\ 
\cdot\sum_j \zeta_{j}\cdot \epsilon_j\cdot\sqrt{\epsilon_j^2-m_\nu^2}\cdot\Theta(\epsilon_j-m_\nu)
\label{eq:betadiff1}
\end{multline}
Equation \ref{eq:betadiff1} contains the square of the nuclear matrix element $ |M_\mathrm{nucl}^2| $, the neutrino kinetic energy $ \epsilon_j = E_0-E-V_j $, the Fermi constant $ G_\mathrm{F} $, the Cabibbo angle $ \Theta_\mathrm{C} $, the electron mass $ m_\mathrm{e} $, and the Fermi function $ F(E,Z' = 2) $.
$ A_\mathrm{s} $ is the normalization of the tritium $ \beta $-decay, and $ N_\mathrm{T} $ denotes the number of tritium atoms in the source multiplied with the solid angle of the setup $ \Delta\Omega/4\pi =(1 - \mathrm{cos}\theta_\mathrm{max})/2 $, with $ \theta_\mathrm{max} = 50.5^{\circ}$, and the detector efficiency of 0.95. The calculation of $ R_\beta(E) $ involves a sum over a molecular final-state distribution (FSD), which describes the probabilities $ \zeta_j $ with which the daughter ion $ ^3\mathrm{HeT}^+ $ is left in  a rotational, vibrational, and/or electronic state with excitation energy $ V_j $ added to the daughter molecule recoil, as described in \cite{KNM1_PRD}. Finally, our calculations include radiative corrections and the thermal Doppler broadening due to the motions of the tritium molecules in the WGTS \cite{Kleesiek_2019}. \newline

\begin{figure}[ht!]
	\includegraphics[width=0.45\textwidth]{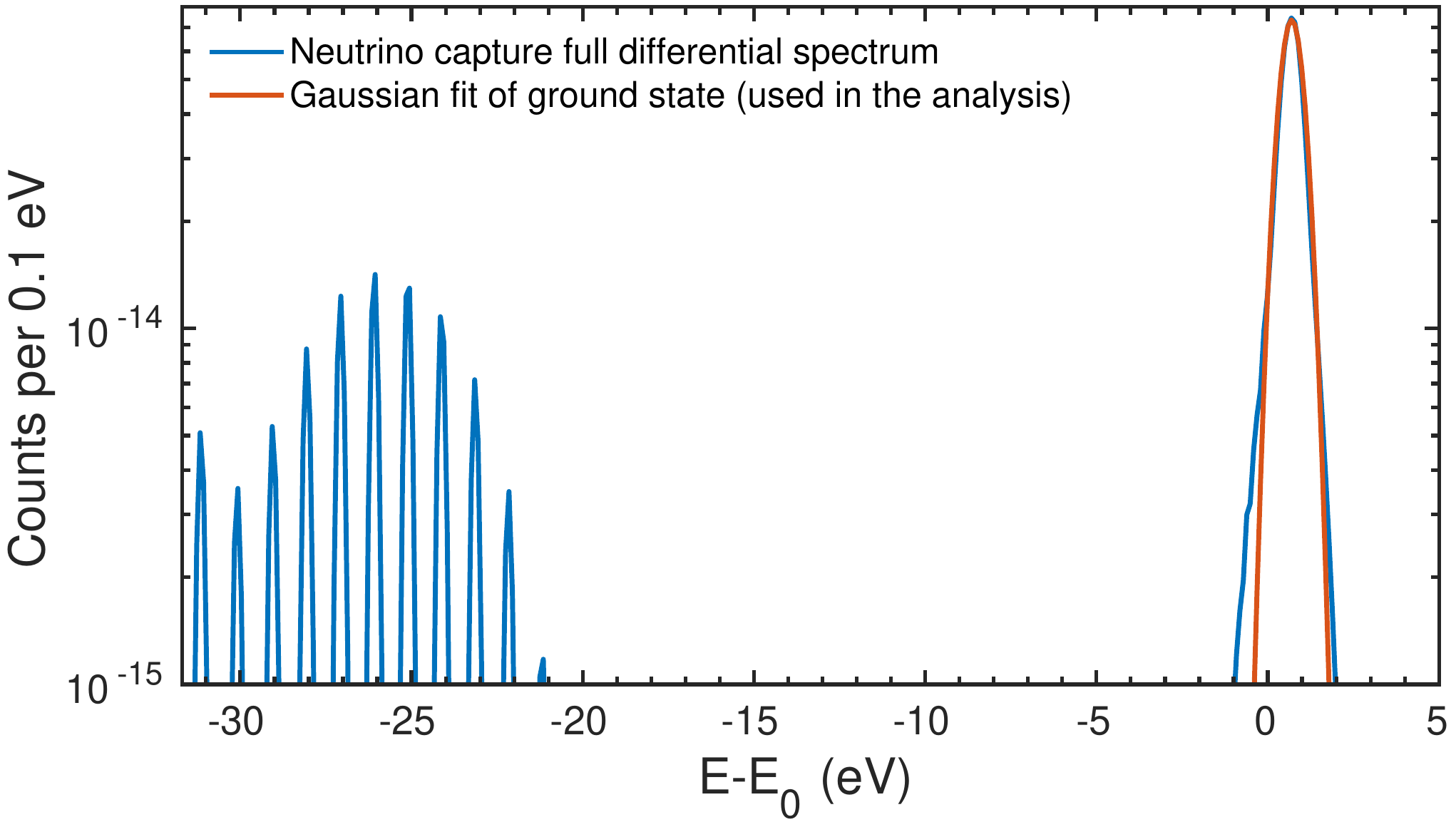}
	\caption{\label{fig:diffspec} Relic neutrino differential signal with arbitrary exposure, full model (blue) and Gaussian simplified model (red) for $\eta$ = 1 and $m_\nu$ = \SI{0.7}{\electronvolt}.}
\end{figure}

{\em Relic neutrino analysis} -- Relic neutrinos have a negligible $\mathcal{O}$(meV) kinetic energy compared to the energy released in the neutrino capture reaction on tritium. The differential neutrino capture spectrum is therefore given by the convolution of a Dirac delta function at $ E_0 + 2m_\nu $ with the FSD. The mean excitation energies of the FSD are given relative to the ro-vibrational
ground state of $^3 \mathrm{HeT}^+$ and include the recoil energy for $^3 \mathrm{HeT}^+$ \cite{PhysRevLett.84.242}.
Electronic excited states are involved in \SI{43}{\percent} of neutrino captures that shift part of the neutrino capture signal well below the endpoint (Fig.~\ref{fig:diffspec}), where it is buried by the prominent $\beta$-decay background. These captures are therefore neglected in what follows. The remaining signal reflects captures involving the molecular ro-vibrational states, which are the same, to a good approximation, for $ \beta$-decay and neutrino capture \cite{PhysRevC.91.035505}. 
We further simplify by modeling the ro-vibrational and the Doppler broadenings by a \SI{0.4}{\electronvolt}-wide Gaussian distribution. Simulations confirm that this proxy has no substantial impact on our results.

%CnuB_DiffSpec.pdf

The neutrino capture differential spectrum is then:
\begin{multline}
	R_{\mathrm{C\nu B}}(E)=\frac{\eta\cdot N_\mathrm{T}\cdot\epsilon_\mathrm{FSD}\cdot R_\mathrm{cap}\cdot T}{\sqrt{2\pi\sigma^2_\mathrm{C\nu B}}}\\
	\cdot\mathrm{exp}\left(-\frac{(E-E_0-m_\nu+\langle E_\mathrm{GS}\rangle)^2}{2\sigma^2_\mathrm{C\nu B}}\right)
	\label{eq:cnubdiff} \, ,
\end{multline}
where $ \eta $ denotes the local relic neutrino overdensity, $\epsilon_\mathrm{FSD} $ the ground-state fraction of the FSD, $ R_\mathrm{cap} = $ \SI{4.2e-25}{yr^{-1}} the relic neutrino capture rate on a single tritium nucleus for Majorana neutrinos \cite{hodak2011beta}, and $ T $ the measurement time. For Dirac neutrinos in the non-relativistic limit, $ R_\mathrm{cap} $ is halved \cite{hodak2011beta}. The final values quoted in this work are given for Majorana neutrinos. 

In the case of molecular tritium, the endpoint of the $\beta$ decay spectrum is broadened by the FSD ground state, and effectively extended by the mean ground state energy $ \langle E_\mathrm{GS}\rangle\approx$\SI{1.7}{eV}, which is equal to the recoil energy of a tritium molecule at the endpoint. This is due to the extended distribution of the FSD ground state, which includes a small fraction of $ \beta $-decay electrons that only loose a minor amount of energy to molecular excitations. This leads to an irreducible background of the relic neutrino signal for neutrino masses smaller than $ \langle E_\mathrm{GS}\rangle/2 = $ \SI{0.85}{\electronvolt}, as the C$\nu$B signal overlaps with the $ \beta$-decay spectrum tail. 
% FSDeffects.pdf
\begin{figure}[ht!]
	\includegraphics[width=0.45\textwidth]{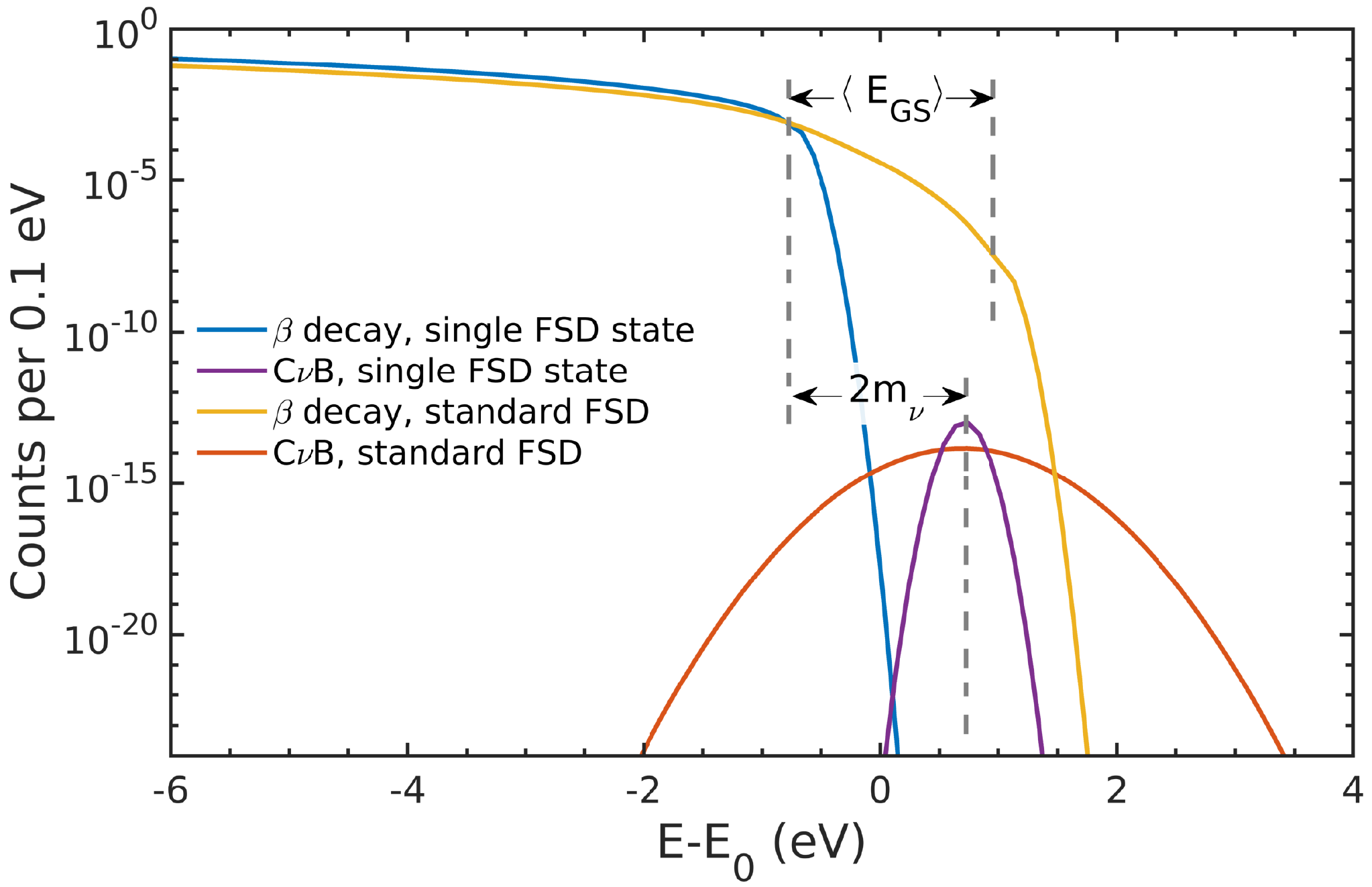}
	\caption{\label{fig:fsdeff}Simulated impact of the molecular effects (FSD) on the $ \beta$-decay and neutrino capture differential spectra of molecular tritium, for $\eta$ = 1 and $m_\nu$ = \SI{0.7}{\electronvolt}  (arbitrary exposure). All spectra also include Doppler broadening.}
\end{figure}
Figure~\ref{fig:fsdeff} illustrates this situation for $\eta = 1$. The $\beta$ spectrum with molecular tritium (yellow) is extended by a high-energy tail compared to the $\beta$ spectrum without FSD smearing (blue), and exceeds the C$\nu$B signal (purple) by about 7 orders of magnitude. This ratio can only be improved by an increased C$\nu$B overdensity. In view of the above considerations, and as discussed in \cite{Cheipesh:2021fmg}, it seems unlikely that any experiment using a molecular tritium source could achieve $ \eta$ limits well below \num{1e6} for neutrino masses below \SI{0.85}{\electronvolt}, making the actual detection of C$\nu$B with molecular tritium virtually impossible. 

Finally, in our analysis we only allow the neutrino mass squared to be positive, for two reasons: First, the average energy of the neutrino capture signal $E_\mathrm{C\nu B} = E_0+2m_\nu-\langle E_\mathrm{GS}\rangle$ depends on the true neutrino mass, being undefined if $m_\nu^2<0$. Second, while negative values of $m_\nu^2$ are called for in neutrino mass analyses to account for a possible rate overshoot near the endpoint (see \cite{KNM1_PRD}), such an overflow is still best explained with a positive (physical) neutrino capture signal.\newline

\begin{figure}[ht!]
	\includegraphics[width=0.45\textwidth]{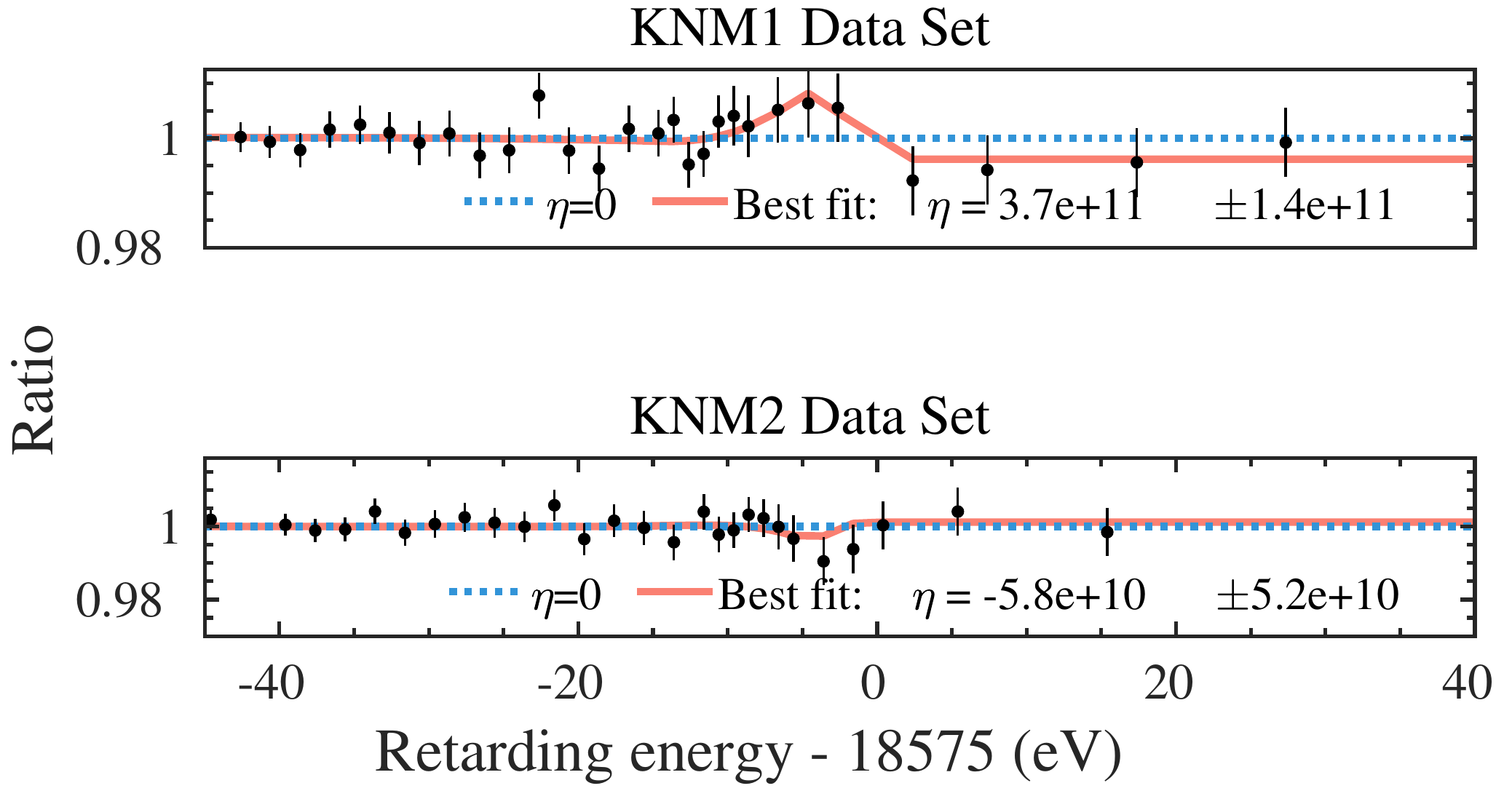}
	\caption{\label{fig:Fits} Spectral ratio of the best fits for relic neutrinos with respect to the null hypothesis, $ \eta $ = 0, top) for the first and bottom) for the second measurement campaigns. The non-zero values of the best fit of $ \eta $ are both consistent with background fluctuations.}
\end{figure}

\begin{figure}[ht!]
   \begin{center}
    	\includegraphics[width=0.45\textwidth]{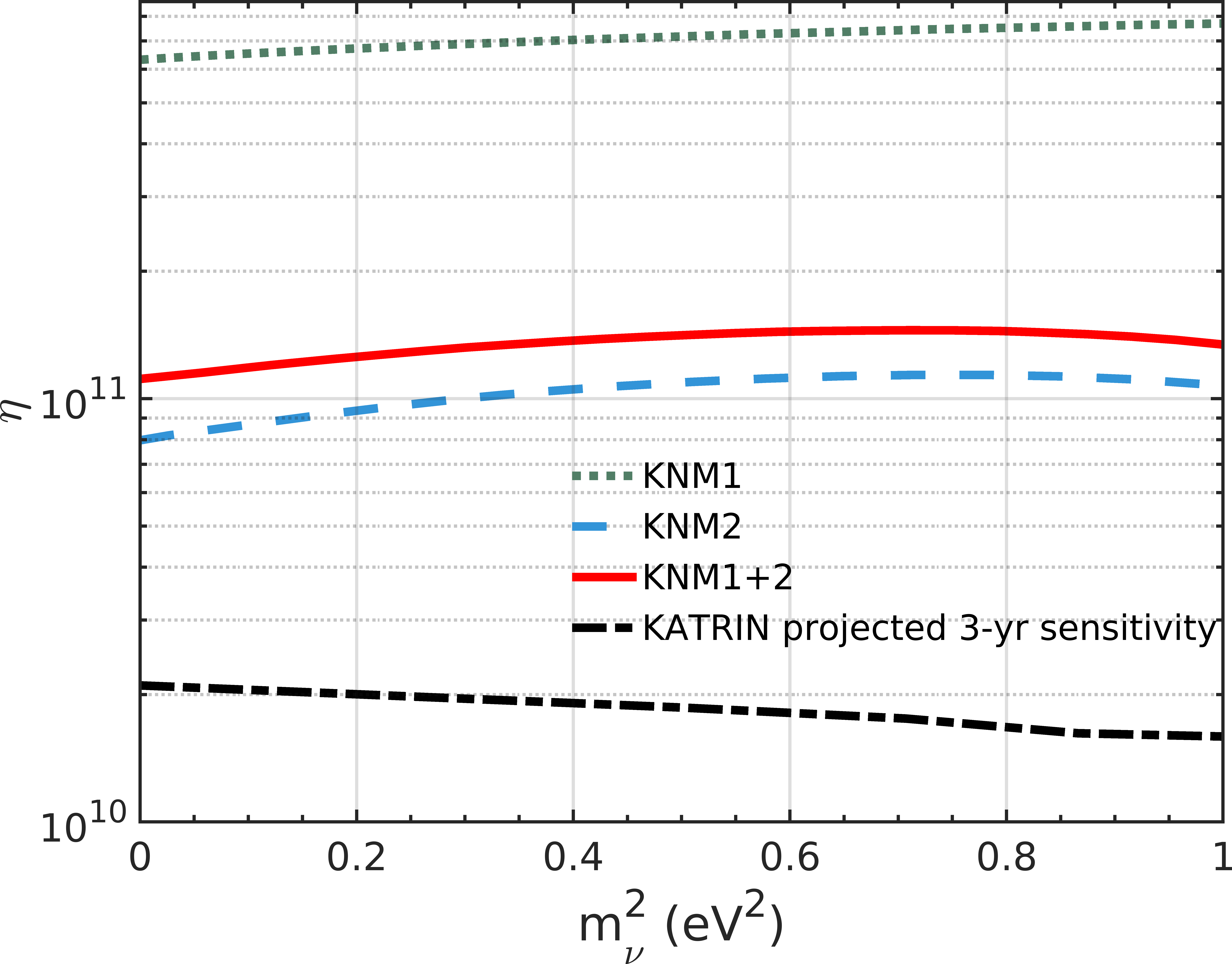}
	\caption{\label{fig:2d} Exclusion contours on the relic neutrino overdensity $\eta$ at 99\% C.L., for both measurement campaigns and their combination, and the projected KATRIN sensitivity for each neutrino mass.}
	\end{center}
\end{figure}

{\em KNM1 results} -- We perform a global fit over the analysis range $ [E_0-\SI{37}{\electronvolt},~E_0+\SI{49}{\electronvolt}] $, treating $  A_\mathrm{s},~E_0,~m_\nu^2,~R_{\mathrm{bg}} $ and $\eta$ as free fit parameters. The covariance matrix approach is applied to propagate systematic uncertainties in the final result in exactly the same way as described in \cite{KNM1_PRD}. The minimization of $\chi^2 $ yields a goodness of fit with a $\chi^2 $ per degree of freedom of 0.81, reflecting a p-value of 0.71. The best fit value for $m_\nu^2=0.8\pm$\SI{0.8}{\electronvolt^2} is consistent with the value obtained in the neutrino mass analysis \cite{KNM1_PRD}. Concerning the main observable $ \eta $, we obtain a best fit value of $(3.7 \pm 1.4)\cdot 10^{11}$ with an uncertainty fully dominated by statistics. This fit is displayed by the red line in the top panel of Fig. \ref{fig:Fits}. 
The $ \left( \Delta\chi^2 \right)_{\mathrm{0H}}$ between the best fit and the null hypothesis ($ \eta\approx0 $) is 3.7. On the basis of 1000 simulated pseudo-experiments, the probability of obtaining $ \eta\geq $\num{3.7e11} if the null hypothesis (0H) is true is 2\%. We therefore conclude that our result does not provide evidence for a relic neutrino signal.
To obtain the $ \eta $ exclusion contour, we first  perform a scan over both $ m_\nu^2 $ and $ \eta $ ranging over [\SI{0}{\electronvolt^2}, \SI{1}{\electronvolt^2}] and [0, \num{8.5e11}] respectively. For each fixed pair of $ m_\nu^2 $ and $ \eta $, we perform the fit of the spectrum  marginalizing over the remaining free fit parameters $  A_\mathrm{s},~E_0 $ and $ R_{\mathrm{bg}} $. When accounting for the positive best fit on the data, the resulting 99\% C.L. exclusion limit shown in Figure~\ref{fig:2d} is in good agreement with our simulations based on the same full analysis performed on a Monte Carlo copy of the data (99\% sensitivity: $\eta<$\num{4.5e11}).  
%As a crosscheck, confidence intervals are also derived from a raster scan over the neutrino mass, calculating the exclusion limit of $ \eta$ for each value of $ m_\nu^2$, which yields similar results. \newline

% BestFits.pdf

{\em KNM2 results} -- This analysis is identical in all respects to the KNM1 case, except for a slightly different set of systematic effects, as described in detail in \cite{aker2021direct}. The fit of both the $\beta$ spectrum and the C$\nu$B signal over the analysis range $[E_0-\SI{40}{\electronvolt},~E_0+\SI{135}{\electronvolt}] $ leads to a $\chi^2 $ per degree of freedom of 1.15, corresponding to a p-value of 0.28. We obtain a best fit value of $ \eta = (-5.8 \pm 5.2)\cdot 10^{10}$, shown by the red line in the bottom panel of Fig. \ref{fig:Fits}, and $m_\nu^2=0.1\pm$\SI{0.3}{\electronvolt^2} in agreement with \cite{aker2021direct}. Again, the uncertainty on $ \eta $ is fully driven by statistics. The value of $ \left( \Delta\chi^2 \right)_{\mathrm{0H}}$ is found to be 1.1. Using 1000 simulated pseudo-experiments mimicking this specific data set, we determine a 5\% probability of obtaining $ \eta\leq $\num{-5.8e10} considering the null hypothesis. As for KNM1, this result is compliant with the null hypothesis and no evidence for a relic neutrino signal arises.
The exclusion limit for the parameter $\eta$ is obtained in the same way as for KNM1 (explained above), and the resulting 99\% C.L. exclusion contour is shown in Fig.~\ref{fig:2d}. The contour is computed with respect to the null hypothesis so as not to benefit from the non-physical best fit related to negative overdensity. The resulting contours are in good agreement with our simulations, both for our main limit, and for the raster scan method also performed for KNM2 (99\% sensitivity: $\eta<$\num{1.8e11}). \newline

{\em Combination of individual results} -- We combine the results of KNM1 and KNM2 to obtain a final result with the full 2019 KATRIN data set. Since the uncertainties of KNM1 and KNM2 are still largely statistically dominated, we can safely presume that the correlations between the two analyses are negligible. The two fit parameters $ m_\nu^2 $ and $ \eta $ are common between the data sets. Since we analyze these two parameters individually for both data sets, we can combine the analyses by adding the $ \chi^2$ surfaces on $m_\nu^2$ and $\eta$ together.
%according to eq. \ref{eq:combi}.
%\begin{equation}
%\chi^2_{\mathrm{combi}} = \chi^2_{\mathrm{KRN1}} + %\chi^2_{\mathrm{KRN2}}
%\label{eq:combi}
%\end{equation}
We obtain a minimal $ \chi^2 $ of 50.7 for 50 degrees of freedom at $ (m_\nu^2,\eta) = (0.02~\mathrm{eV}^2,-2.4\cdot10^{10})$. The resulting 99\% C.L. exclusion contour, provided relative to the null hypothesis, is shown in Fig.~\ref{fig:2d}. The combined upper bound is slightly higher than the result from KNM2 alone, owing to the mild positive fluctuation on $\eta$ found in the KNM1 data set. \newline

% Contours.png

\begin{figure}[ht!]
	\includegraphics[width=0.45\textwidth]{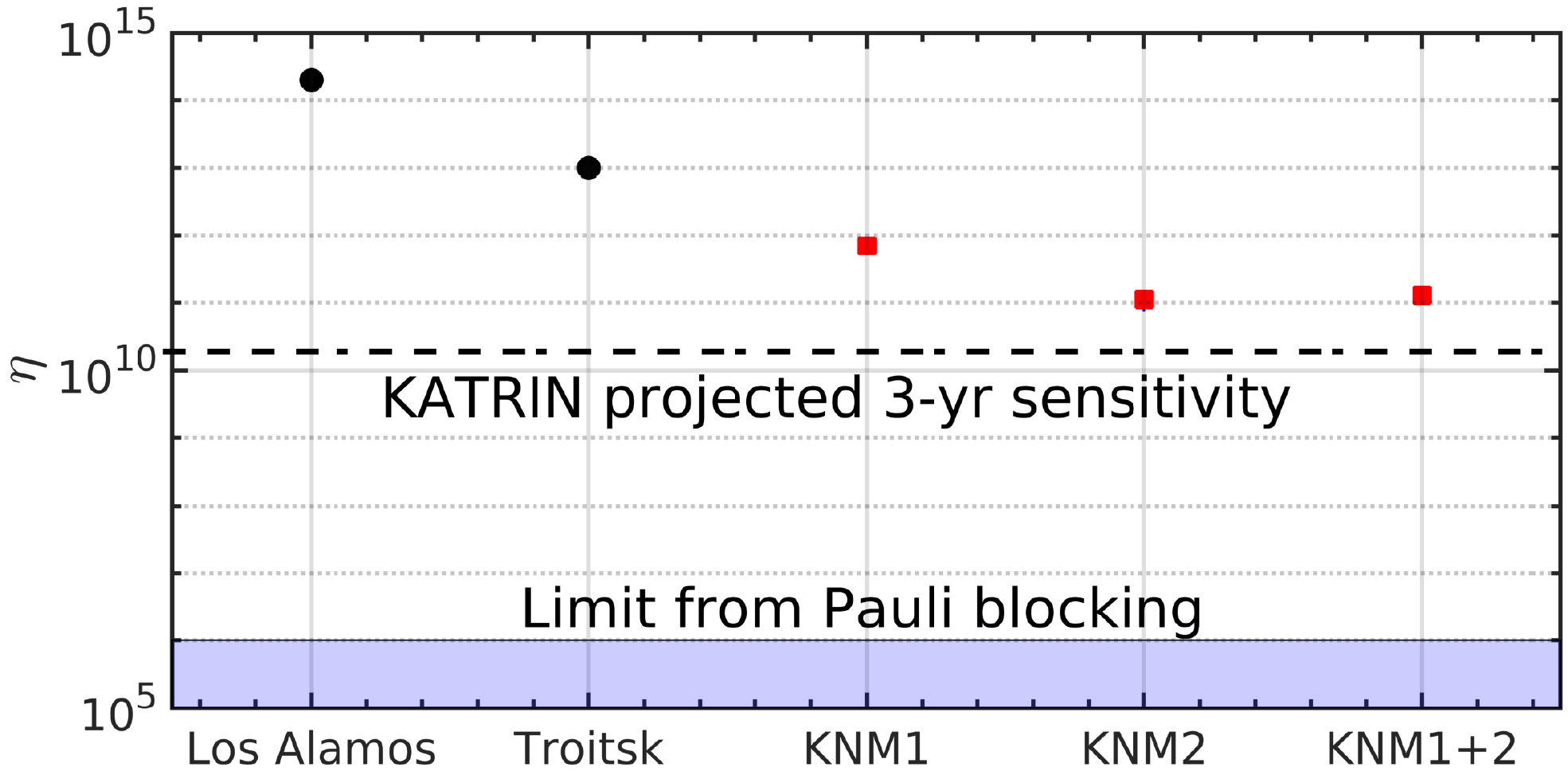}
	\caption{\label{fig:overview} Overview of the new KATRIN limits on the local relic neutrino overdensity, compared to constraints from previous experiments at Los Alamos \cite{Robertson:1991vn} and Troitsk \cite{LOBASHEV1999327}.}
\end{figure}

{\em Final sensitivity forecast} --  Finally, we update the sensitivity of the KATRIN experiment for the search of relic neutrinos based on the current best projection of the experimental settings \cite{TDR_2} and using a background rate $ R_{\mathrm{bg}} $ = \SI{130}{mcps} over all 148 pixels of the detector instead of the \SI{10}{mcps} design value \cite{KDR2004}.
Since the relic neutrino search is very sensitive to any background near the kinematic endpoint, the projected sensitivity is downgraded by a factor of 5 compared to the previous value in \cite{Kaboth_2010}, to reach $\eta < $ \num{1.0e10} (\num{1.4e10}, \num{1.8e10}) at 90\% (95\%, 99\%) C.L., which reproduces the results in \cite{Heizmann2019_1000093536}. Figure~\ref{fig:overview} displays an overview of existing limits on $ \eta $, including our new result and the updated 3-year KATRIN sensitivity, as well as another phenomenological upper limit using the Pauli exclusion principle and the local dark matter density \cite{Read_2014}.\newline

% Overview.pdf

{\em Conclusions and outlook} -- We have searched for a relic neutrino overdensity signal in data from the first two science runs of the KATRIN experiment conducted in 2019. This analysis comprises \num{5.16e6} $ \beta $-decay electrons and \num{0.72e6} background events below $ E_0 $. No significant relic neutrino signal is observed and the parameter $ \eta $ is shown to be less than \num{9.7e10} (\num{1.1e11}, \num{1.3e11}) at 90\% (95\%, 99\%) for neutrino masses below \SI{1}{\electronvolt}, which corresponds to an equivalent number of counts of less than \num{3.2e4} across both data sets. Our result improves on the previous upper limits set by direct kinematics experiments at Los Alamos \cite{Robertson:1991vn} and Troitsk \cite{LOBASHEV1999327}. We have also highlighted that any kinematic experiment searching for the C$\nu$B via neutrino capture using a molecular source suffers from an irreducible background of rare high-energy electrons that are not involved in the energy losses associated with molecular excitations. In the case of tritium, this irreducible background overwhelms the expected C$\nu$B signal considering effective neutrino masses lower than \SI{0.85}{\electronvolt}. 
Finally, KATRIN continues to operate toward the goal of 1000 days of data collected by 2024. The actual elevated background measurements triggered a reassessment of the final sensitivity on the relic neutrino overdensity $ \eta$. We provide an updated sensitivity forecast of $ \eta< $\num{1.0e10} (\num{1.4e10}, \num{1.8e10}) at 90\% (95\%, 99\%) for a background rate of \SI{130}{mcps} across all detector pixels. The PTOLEMY collaboration plans for further improvements in the future \cite{PTOLEMY:2018jst}.  

\begin{acknowledgments}

We acknowledge the support of Helmholtz Association (HGF), Ministry for Education and Research BMBF (05A20PMA, 05A20PX3, 05A20VK3), Helmholtz Alliance for Astroparticle Physics (HAP), the doctoral school KSETA at KIT, and Helmholtz Young Investigator Group (VH-NG-1055), Max Planck Research Group (MaxPlanck@TUM), and Deutsche Forschungsgemeinschaft DFG (Research Training Groups Grants No., GRK 1694 and GRK 2149, Graduate School Grant No. GSC 1085-KSETA, and SFB-1258) in Germany; Ministry of Education, Youth and Sport (CANAM-LM2015056, LTT19005) in the Czech Republic; Ministry of Science and Higher Education of the Russian Federation under contract 075-15-2020-778; and the Department of Energy through grants DE-FG02-97ER41020, DE-FG02-94ER40818, DE-SC0004036, DE-FG02-97ER41033, DE-FG02-97ER41041,  {DE-SC0011091 and DE-SC0019304 and the Federal Prime Agreement DE-AC02-05CH11231} in the United States. This project has received funding from the European Research Council (ERC) under the European Union Horizon 2020 research and innovation programme (grant agreement No. 852845).

%We acknowledge the support of Helmholtz Association, Ministry for Education and Research BMBF (5A17PDA, 05A17PM3, 05A17PX3, 05A17VK2, and 05A17WO3), Helmholtz Alliance for Astroparticle Physics (HAP),  Helmholtz Young Investigator Group (VH-NG-1055), Max Planck Research Group (MaxPlanck@TUM), and Deutsche Forschungsgemeinschaft DFG (Research Training Groups GRK 1694 and GRK 2149, Graduate School GSC 1085 - KSETA, and SFB-1258) in Germany; Ministry of Education, Youth and Sport (CANAM-LM2015056, LTT19005) in the Czech Republic; Ministry of Science and Higher Education of the Russian Federation under contract 075-15-2020-778; and the United States Department of Energy through grants  DE-FG02-97ER41020, DE-FG02-94ER40818, DE-SC0004036, DE-FG02-97ER41033, DE-FG02-97ER41041, DE-AC02-05CH11231, DE-SC0011091, and DE-SC0019304, and the National Energy Research Scientific Computing Center. This project has also received funding from the European Research Council (ERC) under the European Union Horizon 2020 research and innovation program (grant agreement No. 852845).
\end{acknowledgments}  

\bibliographystyle{bst/apsrev4-2}
\bibliography{krn} 

\end{document}